\documentclass[a4paper,11pt]{article}
\pdfoutput=1 % if your are submitting a pdflatex (i.e. if you have
\usepackage{jheppub} % for details on the use of the package, please
\usepackage[T1]{fontenc} % if needed

\usepackage{amsmath,amssymb,tikz}
\usepackage{graphicx}
 \allowdisplaybreaks
\def\be{\begin{equation}}
\def\ee{\end{equation}}
\def\ba#1\ea{\begin{align}#1\end{align}}
\def\bg#1\eg{\begin{gather}#1\end{gather}}
\def\bm#1\em{\begin{multline}#1\end{multline}}
\def\bmd#1\emd{\begin{multlined}#1\end{multlined}}

\def\a{\alpha}
\def\b{\beta}

\def\l{\lambda}

\def\m{\mu}
\def\n{\nu}

\def\la{\label}

\def\({\left(}
\def\){\right)}
\def\[{\left[}
\def\]{\right]}

\def \be {\begin{equation}}
\def \ee {\end{equation}}
\def \ba {\begin{array}}
\def \ea {\end{array}}
\def \bea{\begin{eqnarray}}
\def \eea{\end{eqnarray}}

\def \a {\alpha}
\def \b {\beta}

\def \m {\mu}
\def \n {\nu}

\def \l {\lambda}

\def \la {\leftarrow}
\def \ra {\rightarrow}

\def\bea{\begin{eqnarray}}
\def\eea{\end{eqnarray}}
\newcommand{\eq}[1]{(\ref{#1})}
\newcommand{\bit}{\begin{itemize}}  \newcommand{\eit}{\end{itemize}}
\newcommand{\ben}{\begin{enumerate}}  \newcommand{\een}{\end{enumerate}}

\def\la{\langle}
\def\ra{\rangle}

\def\cA{{\mathscr A}}

\def\one{\mbox{1 \kern-.59em {\rm l}}}

\def\a{\alpha}        
\def\b{\beta}

\def\l{\lambda} 
\def\m{\mu} \def\n{\nu}

\def\cA{{\cal A}}

\def\ba{{\bf a}}

\newcommand{\nn}{{\nonumber}}

 \def\del{\partial}

\def\uno{\mbox{1 \kern-.59em {\rm l}}}

\def\one{1\!\!1\,\,}

\def\bcomment#1{}
\def\IR{\relax{\rm I\kern-.18em R}}

\title{\boldmath  Anomalous Transport in Holographic Boundary Conformal
  Field Theories}

\author[a,b]{Chong-Sun Chu,}
\author[a,c]{Rong-Xin Miao }
\affiliation[a]{Physics Division, National Center for Theoretical Sciences,\\
National Tsing-Hua University, Hsinchu 30013, Taiwan}
\affiliation[b]{ Department of Physics, National Tsing-Hua
  University,
Hsinchu 30013, Taiwan}
\affiliation[c]{ School of Physics and Astronomy, Sun Yat-Sen University, Zhuhai, 519082, China}

\emailAdd{cschu@phys.nthu.edu.tw}
\emailAdd{miaorx@mail.sysu.edu.cn}

\preprint{NCTS-TH/1806}

\abstract{ Recently, it is found that when an external magnetic field
  parallel to the boundary is applied, Weyl anomaly gives rises to a
  new anomalous current transport in the vicinity of the boundary.  At
  the leading order of
  closeness from the boundary,
  the current is determined universally by the
  central charge of the theory. In this paper, we give a holographic
  proof
  for the existence and
  universality for this transport phenomena.
  We show that the current is independent of boundary conditions in four 
dimensions
  while it depends on boundary conditions in other
  dimensions. We also study the backreaction of the bulk Maxwell fields
  on the AdS spacetime and obtain the holographic Weyl anomaly for 5d
  BCFTs in presence of the background field strength.}

\begin{document}

\maketitle
\flushbottom

%c4 \hfill{ NCTS-TH/1802} 
\section{Introduction}

The quantum transportation of charges induced by anomaly is an
interesting phenomenon \cite{review}.  For example, 
the chiral magnetic effect (CME) \cite{Vilenkin:1995um,
Vilenkin:1980fu, Giovannini:1997eg, alekseev, Fukushima:2012vr} refers
to the generation of currents  parallel to an external
magnetic field. And the
chiral vortical effect (CVE) \cite{Kharzeev:2007tn,Erdmenger:2008rm,
  Banerjee:2008th,Son:2009tf,Landsteiner:2011cp,Golkar:2012kb,Jensen:2012kj}
refers to the generation of a current
due to rotational motion in the charged fluid.  Recently, it has
also been pointed out that anomalous
transport also occurs in a conformally flat gravitational spacetime
due to Weyl
anomaly \cite{Chernodub:2016lbo, Chernodub:2017jcp}.
It should be noted that these kinds of anomalous transport
occurs only in a material system where the chemical potentials
are non-vanishing, or in a curved spacetime.  As anomaly is an intrinsic
property of the vacuum, it is interesting to ask if
anomalous current transport could be realized in vacuum without the presence of 
material system.

One of the most well known manifestation of the
quantum nature of the vacuum is the Casimir effect 
\cite{Casimir:1948dh,Plunien:1986ca,Bordag:2001qi}. This occurs since the energy of
the vacuum is sensitive to the change in the boundary condition.
Recently the Casimir effect has been
analyzed in full generality for arbitrary shape of boundary and
for arbitrary spacetime metric, and
universal relations between the Casimir coefficients
and the boundary central charge in a boundary conformal
field theory (BCFT) have been discovered \cite{Miao:2017aba}.
The study was based on a field theory
analysis of the properties of the energy momentum tensor
in the vicinity of boundary. 
In a recent paper \cite{Chu:2018ksb}, we generalized the analysis to boundary system
with  $U(1)$ symmetry and discovered 
a new type of  anomalous current  in the
vicinity of the boundary due to Weyl anomaly
\footnote{
Interestingly,
\cite{Jensen:2017eof} find that there is some evidence that non-Weyl
anomalies such as 't Hooft anomalies are inconsistent with the
existence of a boundary.
}.
In four dimensions, it was found that a current is induced near the boundary
due to the presence of a background field strength
\begin{eqnarray}
  \label{current2}
 J_a  = \frac{4 b_1 F_{a n}}{x}, \quad x \sim 0, \qquad a=0,1,2,3.
\end{eqnarray}
Here $x$ is the geodesic distance from the boundary, $n$ is the direction
of the inward pointing normal, and $b_1$ is
the bulk central charge which appears in the Weyl anomaly
\cite{Duff:1993wm,note}
\be
 \label{A1}
 \mathcal{A}=\int_M \sqrt{g} \Big[ b_1 F_{\m\n}F^{\m\n}+
   \mbox{curvature terms} \Big].
 \ee
For the normalization of the gauge field kinetic term
$ S = -1/(4e^2) \int F^2$, $b_1$ is related to the beta function as 
$b_1 = -\frac{\b(e)}{2 e^3}$. Similar results were also found for higher dimensions
where
the anomalous current is determined universally by the
central charge. 
We remark that 
the induced current \eq{current2} holds for not just boundary conformal
field theories (BCFTs), but also for more general boundary quantum
field theories (BQFTs) which are covariant, gauge invariant, unitary
and renormalizable.
Unlike the previous kinds of anomalous transport, this anomalous
transport occurs in  zero temperature vacuum in flat spacetime without the need
of material support. And it is 
an  intrinsic manifestation of the dependence of the
quantum vacuum on boundary like the Casimir effect.
% vv6
Finally, it should be mentioned that there are boundary contributions to the current density which can exactly cancel the  apparent ``divergence'' in the bulk current (\ref{current2}) at $x=0$ and define a finite total current \cite{Chu:2018ksb}.

BQFTs/BCFTs \cite{Cardy:2004hm,McAvity:1993ue} describe physical 
systems with boundaries.
In recent
years, the field of BCFTs has developed rapidly. In addition to traditional
field theory techniques, see, e.g.
\cite{
  Fursaev:2015wpa,Herzog:2015ioa,Miao:2017aba,Herzog:2017kkj,
Jensen:2017eof, 
Kurkov:2017cdz,Kurkov:2018pjw,Rodriguez-Gomez:2017kxf,Seminara:2017hhh},
holographic models of BCFTs have been developed in 
\cite{Takayanagi:2011zk,Fujita:2011fp,Nozaki:2012qd,Miao:2017gyt,Chu:2017aab,
Astaneh:2017ghi,Miao:2017aba} which allow for non-perturbative analysis of the
boundary systems. Holographic dual of BCFT was originally introduced by
Takayanagi
\cite{Takayanagi:2011zk} by considering a Neumann boundary condition in the bulk
dual manifold. However the tensor type embedding equations of the proposal contain
too many constraints in general
and cannot be solved consistently for general  shape of boundary.
The difficulty was analyzed in 
\cite{Miao:2017gyt,Chu:2017aab} and a 
  consistent model of holographic BCFT
  was found by  replacing the tensor type embedding equation by
  a scalar type embedding equation. More recently we found that
  \cite{Miao:2017aba} the original set of tensor embedding equations can also be
  consistent if one is to allow for a non-FG expansion of the metric in the bulk.

  The models \cite{Miao:2017gyt,Chu:2017aab} and 
  \cite{Miao:2017aba}
  of holographic BCFT
  have been applied to study the one point function of stress tensor.
  Boundary Weyl anomaly 
as well as new universal relations between
the generalized Casimir coefficients and the central charges have been obtained
in both models, and the results agree exactly, except for
a different representation of
the central charges as functions of the holographic BCFT parameter. 
In this paper, we study the anomalous current transport in this two models of
holographic BCFTs,  and show that to the leading
order of closeness to the boundary,
the holographic current is determined universally by the central
charges of the Weyl anomaly.
We also show that the current
is independent of boundary conditions
in four dimensions while it depends on boundary conditions in higher
dimensions. The results agree with those obtained in
\cite{Chu:2018ksb} and generalize to theories that do not necessary admit a
Lagrangian formulation.
That the two proposals of
holographic models yield again the same results for the anomalous current. This
confirms our previous speculation that the two models of holographic BCFT 
correspond to two different kinds of
BCFT which admit different holographic descriptions \cite{Miao:2017aba}.

The paper is organized as follows. 
In section 2, we derive the holographic
current for 4d BCFTs and show that it is
independent of boundary conditions and is 
determined entirely by the bulk
central charge. In section 3,
we generalize our discussions to higher dimensions and show that the
holographic current depends on boundary conditions in higher
dimensions. In section 4, we study the back reactions of Maxwell's fields
to AdS spacetime and derive the holographic Weyl anomaly for 5d BCFTs.
The paper is ended with some future discussions in section 5.

%%%%%%%%%%%%%%%%%%%%%%%%%%%%%%%%%%%

\section{Holographic Current for 4d BCFT}

%%%%%%%%%%%%%%%%%%%%%%%%%%%%%%%%%%%

\begin{figure}[t]
\centering
\includegraphics[width=5cm]{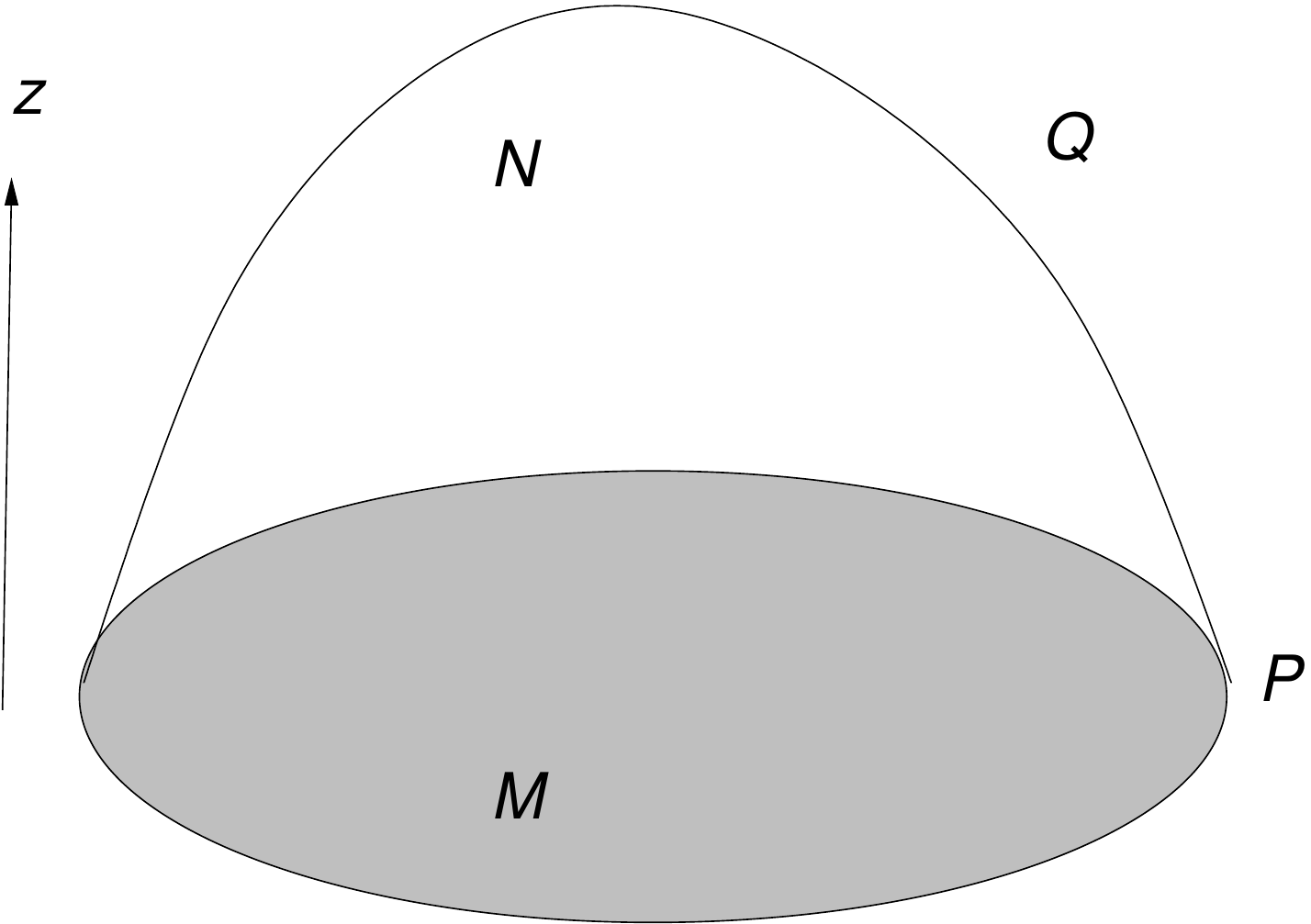}
\caption{BCFT on $M$ and its dual $N$}
\label{MNPQ}
\end{figure}

Consider a boundary conformal field theory (BCFT) defined on a
manifold $M$ with a boundary $P$.
Takayanagi \cite{Takayanagi:2011zk} proposed to extend the $d$
dimensional manifold $M$ to a $(d+1)$ dimensional asymptotically AdS
space $N$ such  that $\partial N= M\cup Q$, where $Q$ is a $d$
dimensional manifold with boundary $\partial Q=\partial
M=P$. See figure \ref{MNPQ}.

To investigate the renormalized current in holographic models
of BCFT,
let us add a $U(1)$ gauge field to the holographic model.
Following \cite{Takayanagi:2011zk,Fujita:2011fp,Nozaki:2012qd},
we consider the following gauge invariant action for holographic BCFT 
 ($16\pi G_N =1$)
\begin{eqnarray}\label{action1}
  I&=&\int_N \sqrt{G} (R-2 \Lambda-\frac{1}{4}\mathcal{F}_{\mu\nu}
    \mathcal{F}^{\mu\nu} ) +2\int_Q \sqrt{\gamma} (K-T).
\end{eqnarray}
Here $\mathcal{F}$ is the bulk field strength which reduces  to $F$
on the boundary $M$. The constant parameter
$T$  can be regarded as the holographic dual
of boundary conditions of BCFT
\cite{Miao:2017gyt,Chu:2017aab}.
The central issue in the construction of the AdS/BCFT
is the determination of the location of $Q$ in the
bulk.
Takayanagi \cite{Takayanagi:2011zk} proposed to impose the Neumann boundary
condition (NBC) on $Q$. In our case this means we have on $Q$:
\bea
&K_{\alpha\beta}-(K-T)\gamma_{\alpha\beta} =0, \label{NBC-g}\\
&\mathcal{F}_{\mu\nu}n_Q^{\mu}\Pi^{\nu}_{\ \alpha}=0, \label{NBC-A}
\eea
where $n_Q$ is the inward-pointing normal vector on $Q$, and $\Pi$ is
the projection operator which gives the vector field and metric on $Q$: 
$\bar{A}_{\alpha}=\Pi^{\nu}_{\ \alpha}\mathcal{A}_{\mu}$ and
$\gamma_{\alpha\beta}=\Pi^{\mu}_{\ \alpha}\Pi^{\nu}_{\ \beta}G_{\mu\nu}$.
As discussed in \cite{Miao:2017gyt,Chu:2017aab},
the tensor embedding equations  \eq{NBC-g} admit
no solution in general if the  bulk metric is assumed to admit a FG expansion.
On the other hand, \eq{NBC-g} becomes consistent 
if a non-FG expanded bulk metric is adopted. 
Another possible holographic model for BCFT  \cite{Miao:2017gyt,Chu:2017aab}
is to  consider a mixed boundary
condition on the metric whereby one obtains the scalar embedding equation
\be
K = \frac{d}{d-1} T, \label{NBC1-g}
\ee
together with \eq{NBC-A}. This model is also consistent.
Both models yield the same
consistent results for the  Weyl anomaly 
as well as giving the same universal relations between
the generalized Casimir coefficients and the central charges,
the later of which are
however parametrized
differently in terms of  $T$ for the two holographic models.
In the following, we will consider both proposals of holographic BCFT.
Quite amazingly, we find again that both
models give the same results for the induced current when the result
is expressed in terms of the central charges.
This is quite remarkable and is a reconfirmation of the earlier
indication that the two proposals of
holographic models  corresponds to two different kinds of
BCFT \cite{Miao:2017aba}.

Now back to our case.
For simplicity, let us consider the case of a  flat half space $x\geq 0$.
The bulk metric reads
\begin{equation}\label{AdSmetric}
ds^2=\frac{dz^2+dx^2+\delta_{ab}dy^ady^b}{z^2}.
\end{equation}
In this case, \eq{NBC-g} reduces to \eq{NBC1-g}, and $Q$ is given by
\cite{Takayanagi:2011zk}
\begin{equation}\label{Q}
\ x=-\sinh \rho\ z,
\end{equation}
where we have re-parametrized $T=3 \tanh \rho$. As for
the solution for the vector field,  due to the
planar symmetry of the boundary, we consider $A_{\mu}$ that depends
only on the coordinates $z$ and $x$. The Maxwell equations
$\nabla_{\mu}\mathcal{F}^{\mu\nu}=0$ can be solved with
$\mathcal{A}_z=\mathcal{A}_z(z)$, $\mathcal{A}_x=\mathcal{A}_x(x)$ and
$\cA_a$ satisfying,
\begin{eqnarray}\label{EOMvector}
z \del_x^2 \mathcal{A}_a -
\del_z \mathcal{A}_a+z \del_z^2 \mathcal{A}_a=0.
\end{eqnarray}
One can solve the above equation by separation of
variables $\mathcal{A}_a(z,x)=Z(z)X(x)$, and then substitute the general solutions
to (\ref{NBC-A}) to obtain the solution by brute force.  However there is a
quicker trick.
Inspired by similar considerations in \cite{Miao:2017aba},  
let us take the following ansatz for the vector field
\begin{eqnarray}\label{vectoransatz}
  \mathcal{A}_a=A^{(0)}_a+  x f_1(\frac{z}{x}) A^{(1)}_a
  + x^2 f_2(\frac{z}{x}) A^{(2)}_a+
  \cdots,
\end{eqnarray}
where we set $f_i (0)=1$ so that $\mathcal{A}_a$ reduce
to the guage field $A_a$  at the AdS boundary $z=0$. Here 
$A^{(i)}$ are the expansion coefficients of $A_a$ about the boundary: 
\begin{eqnarray}\label{vectorgauge}
A_a=A^{(0)}_a+ xA^{(1)}_a+ \cdots .
\end{eqnarray}
In particular, $A_a^{(1)}$ is given by the field strength at the boundary:
\be
\label{AF}
A_a^{(1)} = F_{xa} = F_{na}.
\ee
Note that
in the derivative expansions we have $O(A^{(i)}) \sim O(\partial)^i$.
Substituting (\ref{vectoransatz}) into (\ref{EOMvector}) we get 
\begin{eqnarray}
s (s^2+1) f_1''(s)-f_1'(s)=0, 
\end{eqnarray}
at the linear order $O(\partial)$. Recall that $f_1(0)=1$,
we have the solution $f_1(s)=1-c_1+c_1 \sqrt{1+s^2}$, and
(\ref{vectoransatz}) reads
\begin{eqnarray}\label{vectoransatz1}
\mathcal{A}_a=A^{(0)}_a+  \left( (1-c_1) x+ c_1\sqrt{x^2+z^2} \right) A^{(1)}_a,
\end{eqnarray}
where we have ignored the higher order terms since they are irrelevant
to the current (\ref{current2}) of order $O(\partial)$, or equivalently,  $O(F)$.
Note also that we have analytic continuated $ x \sqrt{1+\frac{z^2}{x^2}}$ to
$ \sqrt{x^2+z^2}$ in order to get smooth solution at $x=0$. 
Imposing the boundary condition (\ref{NBC-A}) on $Q$, we get 
$c_1=1$. One can check directly that the solution $\mathcal{A}_z=\mathcal{A}_z(z),
\mathcal{A}_x=\mathcal{A}_x(x)$ and 
\begin{eqnarray}\label{vectors}
\mathcal{A}_a=A^{(0)}_a+ A^{(1)}_a \sqrt{x^2+z^2}
\end{eqnarray} 
is  indeed an exact solution to the
Maxwell equations and the boundary condition (\ref{NBC-A}) in $AdS$. 

From the gravitational action (\ref{action1}), we can derive the holographic
current as \cite{Gynther:2010ed}
\begin{eqnarray}\label{holocurrent0}
  \la J^a \ra=\lim_{z\to 0}\frac{\delta I}{\delta A_a}
  =\lim_{z\to 0}\sqrt{G}\mathcal{F}^{za}
\end{eqnarray}
Substituting the solutions (\ref{AdSmetric}), (\ref{vectors})
into (\ref{holocurrent0}), we obtain
\begin{eqnarray}\label{holocurrent}
\la J_a \ra=\partial_z^2\mathcal{A}_a|_{z=0}= -\frac{F_{an} }{x} +O(1),
\end{eqnarray}
where we have used \eq{AF}. 
On the other hand, the holographic Weyl anomaly of (\ref{action1}) is
obtained in \cite{Genolini:2016ecx} with the central charge given by
%$b_1=-\frac{1}{4}.$
\begin{eqnarray}\label{holocharge}
b_1=-\frac{1}{4}.
\end{eqnarray}
Now it is clear that the holographic BCFT satisfies the universal law
of current (\ref{current2}).  
It is remarkable that current (\ref{holocurrent}) is independent of the
parameter $T$ , which shows that
near-boundary current for 4d BCFT is indeed independent of boundary conditions.

%%%%%%%%%%%%%%%%%%%%%%%%%%%%%%%%%%%
\section{Holographic Current in Higher Dimensions}
%%%%%%%%%%%%%%%%%%%%%%%%%%%%%%%%%%%

In this section, we study the holographic current for BCFTs in higher
dimensions. We verify that the leading term of the current
is determined  universally by the central charge of Weyl
anomaly. However unlike the 4d case, the current depends on boundary
conditions in higher dimensions.

We start with the
gravitational action 
\begin{eqnarray}\label{actionhigher}
  I=\int_N \sqrt{G} \Big(
  R-2 \Lambda-\frac{1}{4}\mathcal{F}_{\mu\nu}\mathcal{F}^{\mu\nu}
  \Big)
  +2\int_Q \sqrt{\gamma} (K-T).
\end{eqnarray}
We work in AdS spacetime (\ref{AdSmetric}) with the bulk boundary $Q$
given by (\ref{Q}). Focus on the leading order of current, we can
ignore the back reactions of the Maxwell's fields to spacetime and the
location of the bulk boundary. 
For  plane boundary,  $A_{\mu}$ depends on
only the coordinates $z$ and $x$. For general dimension $d$,
the Maxwell's equations can be solved with
$\mathcal{A}_z=\mathcal{A}_z(z)$,
$\mathcal{A}_x=\mathcal{A}_x(x)$ and
\begin{eqnarray}\label{EOMvectorhigher}
  z \del_x^2 \mathcal{A}_a -(d-3)\del_z \mathcal{A}_a
  +z \del_z^2 \mathcal{A}_a =0.
\end{eqnarray}
Similarly, the boundary condition \eq{NBC-A} becomes
\begin{equation}\label{vectorBCs1}
 ( \del_x \mathcal{A}_a+\sinh\rho
  \del_z \mathcal{A}_a ) \Big|_{x=-z \sinh \rho } =0.
\end{equation}
As before, we consider the ansatz for the gauge field
\begin{eqnarray}\label{vectoransatz-5d}
\mathcal{A}_a=A^{(0)}_a+ A^{(1)}_a x f(\frac{z}{x}),
\end{eqnarray}
where $f(0)=1$ and $A^{(i)}_a$ are constants. The Maxwell's equations
(\ref{EOMvectorhigher}) becomes
\begin{eqnarray}
s (s^2+1) f''(s)-(d-3)f'(s)=0. 
\end{eqnarray}
It  has the general solution
\begin{eqnarray}\label{vectorsolutionhigher}
  f(s)=1+\alpha_d  \frac{s^{d-2} \, _2F_1\left(\frac{d-3}{2},
    \frac{d-2}{2};\frac{d}{2};-s^2\right)}{d-2},
\end{eqnarray}
where $\alpha_d$ is an integration constant. It should be mentioned that,
in order to get regular solutions at $x=0$, suitable analytic
continuation of the hypergeometric function should be taken when one
express the above solutions in terms of the coordinates $z$ and $x$.
With the ansatz (\ref{vectoransatz-5d}), the boundary condition
(\ref{vectorBCs1}) is simplified to
\begin{equation}\label{vectorBCs2}
\cosh\rho \coth\rho f'(-\text{csch}\rho)+f(-\text{csch}\rho)=0.
\end{equation}
Substituting (\ref{vectorsolutionhigher}) into (\ref{vectorBCs2}), we get
the integration constant
\begin{equation}\label{integralconstant}
  \alpha_d =\frac{(2-d) \text{csch}^3\rho (-\coth\rho)^d}{\text{csch}\rho \,
    _2F_1\left(\frac{d-3}{2},\frac{d-2}{2};\frac{d}{2};-\text{csch}^2\rho\right)
    (\coth \rho \ \text{csch}\rho)^d
    +(d-2) \cosh\rho \coth ^4\rho (-\text{csch}\rho)^d}.
\end{equation}
Notice that 
suitable analytic
continuation of the hypergeometric function should be taken
in order to get smooth function
at $\rho=0$. For example, we have for $d=4,5$, 
\begin{eqnarray}\label{integralconstant1}
  \alpha_4=1,\qquad
  \alpha_5=\frac{2}{\pi+4 \tan ^{-1}\left(\tanh \left(\frac{\rho }{2}
    \right)\right) }.
\end{eqnarray}

Now we are ready to derive the holographic current.
Substituting \eq{AdSmetric}, \eq{vectoransatz-5d} and
\eq{vectorsolutionhigher} into \eq{holocurrent0}, we obtain
\begin{eqnarray}\label{holocurrenthigher}
  \la J_a \ra=\lim_{z\to 0}\frac{\partial_z\mathcal{A}_a}{z^{d-3}}=
  -\alpha_d\frac{F_{an} }{x^{d-3}} +O(\frac{1 }{x^{d-4}}),
\end{eqnarray}
which takes the correct near-boundary behavior \cite{Chu:2018ksb}
and agrees with the 4d
result (\ref{holocurrent}). Note that $\alpha_d$ depends on $\rho$ for
$d\ge 5$, which means that the near-boundary current depends on the
boundary conditions in higher dimensions. According to
\cite{Chu:2018ksb}, $\alpha_d$ is related to the central charge of
Weyl anomaly.  In the next section, we will show that, similar to the
4d BCFTs, this is also the case for 5d BCFTs with gravity duals.

\section{Holographic Weyl Anomaly}

In this section, we calculate the holographic Weyl anomaly of 5d
BCFTs. For our purpose, it is sufficient
to focus on the flat space with a plane
boundary. Then all the curvatures and extrinsic curvatures vanish and
only the field strength contribute to Weyl anomaly. In this case the Weyl anomaly
takes the form
\begin{eqnarray}\label{5dWeylanomaly}
\mathcal{A}=\int_{\partial M}dx^4 \sqrt{h} [ b_1 F_{na}F^{na}+b_2 F_{ab}F^{ab}],
\end{eqnarray}
where $b_1$ and $b_2$ are boundary central charges, which depend on BCs. 
In \cite{Chu:2018ksb} we have derived the relation between the current coefficient
and the boundary central charge
\begin{eqnarray}\label{5drelation}
\alpha_5=2b_1
\end{eqnarray}
that holds  universally for any BCFT with a path integral formulation.
Using our holographic Weyl anomaly derived below,
we prove that the universal relation (\ref{5drelation}) is verified
for any holographic
BCFT.

According to \cite{Henningson:1998gx}, holographic Weyl anomaly can be
obtained as the UV logarithmic divergent terms of the gravitational
action (\ref{actionhigher}). Since the Weyl anomaly
(\ref{5dWeylanomaly}) is of order $O(F^2)$, we need to solve for the
metric up to  order $O(F^2)$
and the gauge field up to  order $O(F)$. This
means that we have to take into account of the back reaction of the Maxwell's
fields to the spacetime metric.
The solutions of the Maxwell fields up to order $O(F)$ has been
obtained in section 3.
Without loss of generality, we set all components of the gauge field to be zero
except $\mathcal{A}_{y_1}$:
\begin{eqnarray}\label{vector5d}
\mathcal{A}_{\mu}=(0,0, F x f(\frac{z}{x}),0,0).
\end{eqnarray}
Here we have denoted $F :=F_{n y_1}$ and 
$f(s)$ is given by  \eq{vectorsolutionhigher}, \eq{integralconstant1}. 
Inspired by \cite{Miao:2017aba}, we consider the following ansatz of
metric for $x>0$
\begin{eqnarray}\label{bulkmetrick5dapp}
 ds^2=\frac{1}{z^2}\Big{[} dz^2+ \big(1+F^2x^4
   X(\frac{z}{x})\big)dx^2 +\big(\delta_{ab} +F^2 x^4 Q_{ab}(\frac{z}{x})
   \big)dy^a dy^b\Big{]}+O(F^3),
\end{eqnarray}
where $Q_{ab}(s)=\text{diag}\left(Q_1(s),Q_2(s),Q_3(s),Q_4(s)\right)$.
The solutions for $x<0$ can be obtained by analytic continuation.
The ansatz (\ref{bulkmetrick5dapp}) works well in odd dimensions.
For even-dimensional BCFTs, we need to add extra
terms to (\ref{bulkmetrick5dapp}) due to the presence of bulk Weyl anomaly. We
consider
\begin{eqnarray}\label{constant5dapp}
X(0)=0,\quad  Q_{ab}(0)=0, 
\end{eqnarray}
so that the BCFT lives in a flat space. 
Solving Einstein-Maxwell equations up to order $O(F^2)$
\begin{eqnarray}\label{EinsteinMaxwellapp}
  R_{\mu\nu}-\frac{R-2\Lambda}{2}G_{\mu\nu}=\frac{1}{2}\left(
  \mathcal{F}_{\mu \rho} \mathcal{F}_{\nu}^{\ \rho}-
  \frac{1}{4}\mathcal{F}_{\sigma \rho} \mathcal{F}^{\sigma \rho}G_{\mu\nu}   \right),
\end{eqnarray}
we obtain
\begin{eqnarray}\label{solutions5dapp}
  96X(s)&=&15 s^4+\alpha _5 \left(34 s^3+6 s\right)
  +\alpha _5^2 \left(\left(60 s^2+46\right) \log \left(s^2+1\right)-s^2
  \left(21 s^2+43\right)\right)\nonumber\\
  &&-2 \alpha _5 \left(\alpha _5 s \left(17 s^2+3\right)+3
  \left(5 s^4+6 s^2+1\right)\right)g(s) \nonumber\\
&&+3 \alpha _5^2 \left(5 s^4+6 s^2+1\right) g(s)^2,\nonumber\\
  384Q_a(s)&=& 
  12 s \left(5 s^3-4 \lambda_a  \left(5 s^2+3\right)\right)\nn\\
  && -6 \alpha _5 s \left(13 s^2+3\right)  
  +\alpha _5^2  \left(-96 s^4-109 s^2+40 \log \left(s^2+1\right)\right)
  \nn\\
  &&+2  \left(\alpha _5^2 \left(139 s^2+69\right) s +72 \lambda_a
  \left(s^2+1\right)^2+\alpha _5 \left(-39 s^4+42 s^2+9\right)\right)g(s)\nonumber\\
  &&
  -3 \alpha_5^2 \left(7 s^4+54 s^2+23\right) g(s)^2,  \qquad a =1,2,3, 
  \nonumber\\
384Q_4(s)&=&-12 s \left(3s^3 -4 
\left(\lambda _1+\lambda _2+\lambda _3\right) (5 s^2+3) \right) \nn\\
&&  -\alpha _5 s(38  s^2-6)
+ \alpha _5^2 \left(48 s^4-s^2+40 \log \left(s^2+1\right)\right)
\nonumber\\
  &&+\left(\a_5 (90 s^4+36 s^2-6)
  -2 \alpha _5^2 \left(41 s^2+39\right) s-144
  \left(\lambda _1+\lambda _2+\lambda _3\right) \left(s^2+1\right)^2\right) g(s)
  \nn\\
 && 
+\alpha _5^2 \left(-9 s^4+54 s^2+39\right) g(s)^2,
\end{eqnarray}
where $\lambda_1,\lambda_2,\lambda_3$ are integration constants,
$s=z/x$ and $g(s)=\tan^{-1}(s)$ for $x>0$. In order to get continuous
solutions at $x=0$, one must perform suitable analytic continuation for $g(s)$.
In this way, we get $g(x,z) =  \frac{\pi }{2}-2
\tan^{-1}\left(x/(z+\sqrt{z^2+x^2})\right)$. 

% vv6
It should be mentioned that the back-reacted spacetime (\ref{bulkmetrick5dapp},\ref{solutions5dapp}) is well-defined and has no physical divergence at $x=0$. To see this, let us calculate the following geometric invariants. We have
\begin{eqnarray}\label{geometricinvariants1}
&&\lim_{x\to 0}R=-30+\frac{z^4}{16} \left(\left(4+\pi ^2\right) \alpha _5^2-4 \pi  \alpha _5+4\right)  F^2+O(F^3)\\
&&\lim_{x\to 0}R_{\mu\nu}R^{\mu\nu}=150- \frac{5z^4}{8}\left(\left(4+\pi ^2\right) \alpha _5^2-4 \pi  \alpha _5+4\right)  F^2+O(F^3) \label{geometricinvariants2}\\
&&\lim_{x\to 0}R_{\mu\nu\rho\sigma}R^{\mu\nu\rho\sigma}=60-\frac{z^4}{4} \left(\left(4+\pi ^2\right) \alpha _5^2-4 \pi  \alpha _5+4\right) F^2+O(F^3), \label{geometricinvariants3}
\end{eqnarray}
which are indeed finite at $x=0$.  Note that the perturbation solutions work well only for small $F$. Strictly speaking, one can only define `small' for a dimensionless number. Thus by `small' we actually means $F x^2 \ll 1$ and $F z^2 \ll 1$ (Note that $x$ and $z$ can be large but $F x^2$ and $F z^2$ should keep small). From (\ref{geometricinvariants1},\ref{geometricinvariants2},\ref{geometricinvariants3}), it is clear that the back reactions to spacetime are not only finite and also small. Recall that we are interested in the near-boundary current. According to \cite{Chu:2018ksb}, by `near' it means $x \ll 1/\sqrt{F}$. Thus the perturbation solutions with $F x^2 \ll 1$ are sufficient for our purpose. Finally, we want to mention that the back-reacted spacetime (\ref{bulkmetrick5dapp},\ref{solutions5dapp}) is asymptotically AdS in the sence that
\begin{eqnarray}\label{asymptoticallyAdS}
\lim_{z\to 0}R^{\mu\nu}_{\ \ \rho\sigma}=-(\delta^{\mu}_{\rho}\delta^{\nu}_{\sigma}-\delta^{\mu}_{\sigma}\delta^{\nu}_{\nu})+O(F^3),
\end{eqnarray}
where we have set the AdS radius $l=1$.

Following \cite{Miao:2017aba}, we assume embedding function of $Q$ takes the form
\begin{eqnarray}\label{Q5dapp}
x=-\sinh(\rho) z+\lambda_4 F^2 z^5+O(F^3)
\end{eqnarray}
where $\lambda_4$ is a function of $\rho$. This is to be solved using either of
the boundary conditions \eq{NBC-g} or \eq{NBC1-g}, and define  consistent model
of holographic BCFT respectively.
Let us first 
consider the boundary condition \eq{NBC-g} on $Q$ \cite{Takayanagi:2011zk}
\begin{equation}\label{NBC5dapp}
K_{\alpha\beta}-(K-T)\gamma_{\alpha\beta}=0.
\end{equation}
In this case, the intergation constants are fixed by the
extra equations in \eq{NBC5dapp}: 
\begin{eqnarray}\label{solutionconstant5dapp}
  \lambda_1&=& \frac{1}{6} \Big(\alpha _5 \left(2-3 \alpha _5 p(\rho )\right)
  +3 \sinh (\rho ) \left(\alpha _5 p(\rho )+1\right){}^2\Big),\nonumber\\
  \lambda_2&=&\lambda_3=\frac{1}{6} \left(\alpha _5 \left(3 \alpha _5 p(\rho )
  +2\right)-3 \sinh (\rho ) \left(\alpha _5 p(\rho )+1\right){}^2\right),\nonumber\\
  \lambda_4&=&\frac{\alpha _5 (-40 \cosh (2 \rho )-15 \cosh (4 \rho )+87)
    +12 (6 \sinh (\rho )+\sinh (3 \rho ))}{3840}\nonumber\\
  &&+\frac{\alpha _5^2 \left(17 \sinh (\rho )-71 \sinh (3 \rho )
    +4 \sinh ^3(\rho ) (43 \cosh (2 \rho )+57)
    \log \left(\coth ^2(\rho )\right)\right)}{3840}\nonumber\\
  &&-\frac{p(\rho ) \left(\alpha _5 \left(\alpha _5
    (40 \cosh (2 \rho )+15 \cosh (4 \rho )-87)-6 (21 \sinh (\rho )
    +5 \sinh (3 \rho )) \cosh ^2(\rho )\right)\right)}{3840}\nonumber\\
  &&+\frac{\alpha _5^2 p(\rho )^2 (21 \sinh (\rho )+5 \sinh (3 \rho ))
    \cosh ^2(\rho )}{1280},
\end{eqnarray}
where $p(\rho):=-2 \tan ^{-1}\left(\frac{\sinh (\rho )}{\cosh (\rho
  )+1}\right)-\frac{\pi }{2}$ and $\alpha_5$ is given by
(\ref{integralconstant1}).  It is remarkable that the
\eq{NBC5dapp} determines not only the integration constants of the
metric but also the location of bulk boundary $Q$. That is because we
only need one equation to fix the codimension one surface $Q$, however
there are many extra equations in the BC (\ref{NBC5dapp}). The extra
equations help to fix the bulk metric in addition to the location of
$Q$.

Next let us consider the boundary condition \eq{NBC1-g}.
One can solve for $\l_4$ by
substituting the embedding function
(\ref{Q5dapp}) into \eq{NBC1-g}.
It turns out that 
$\lambda_4$ is unchanged and is given by exactly the same expression
(\ref{solutionconstant5dapp}) with $\alpha_5$ given by
(\ref{integralconstant1}). Note that unlike the BC
\eq{NBC-g}, the BC \eq{NBC1-g} does not fix the integration
constants $(\lambda_1,\lambda_2,\lambda_3)$ of the bulk metric
(\ref{solutions5dapp}). It is remarkable that, as we will show below,
the Maxwell field strength part of the Weyl anomaly is independent of these
parameters
$(\lambda_1,\lambda_2,\lambda_3)$.

Now we have worked out the bulk metric and the embedding function of $Q$
up to  order $O(F^2)$. Let us go on to calculate the holographic Weyl
anomaly.
We will keep the parameters $(\lambda_1,\lambda_2,\lambda_3)$ free 
so that our following discussion apply to
both the BCs \eq{NBC-g} and \eq{NBC1-g}. By
using the EOM (\ref{EinsteinMaxwellapp}) and NBC \eq{NBC1-g}, we
can rewrite the on-shell gravitational action (\ref{actionhigher}) as
\begin{eqnarray}\label{action5dapp}
  I=\int_N \sqrt{G} [-10-\frac{1}{8}\mathcal{F}_{\mu\nu}\mathcal{F}^{\mu\nu} ]
  +2\int_Q \sqrt{\gamma} \tanh \rho.
\end{eqnarray}
To get the holographic Weyl anomaly, we need to do the integration
along $x$ and $z$, and then select the UV logarithmic divergent terms.
We divide the integration region into two parts: region I is defined
by $( z \ge 0, x \ge 0)$ and region II is defined by the complement of
region I. Let us first study the integration in region I, where only
the bulk action in (\ref{action5dapp}) contributes. Integrating along
$z$ and then expanding the result in small $z=\epsilon_z$, we obtain
\begin{eqnarray}\label{actionregionI}
  I_1&=&\int_{\epsilon_x} dx [\frac{2\alpha_5}{3x} F_{na}F^{na}
    -\frac{13}{32 \epsilon_z}F_{na}F^{na} -\frac{2}{\epsilon_z^5}+O(1)]\nonumber\\
&=&\log(\frac{1}{\epsilon_x}) \frac{2\alpha_5}{3} F_{na}F^{na} + \cdots .
\end{eqnarray}
Next let us consider the integration in region II. In this case, both the bulk
action and boundary action in (\ref{action5dapp}) contribute. For the
bulk action, we first do the integral along $x$, which yields a
boundary term on $Q$. Note that since only the UV logarithmic
divergent terms are related to Weyl anomaly, we keep only the lower
limit of the integral of $x$. Adding the boundary term from bulk
integral to the boundary action in (\ref{action5dapp}), we obtain
\begin{eqnarray}\label{actionregionII}
  I_2&=& \int_{\epsilon_z} dz [-\frac{\alpha_5}{6z}F_{na}F^{na}
    +\frac{2 \sinh (\rho )}{z^5}+O(1)]\nonumber\\
&=&\log(\frac{1}{\epsilon_z}) \frac{-\alpha_5}{6} F_{na}F^{na} + \cdots .
\end{eqnarray}
Note that the results \eq{actionregionI} and \eq{actionregionII} are
independent of $\l_1, \l_2, \l_3$.
Note also 
that $\log(\frac{1}{\epsilon_x})$ and $\log(\frac{1}{\epsilon_z})$
are counted as the same divergence since they differ only by a finite term
$\log l $ for $\epsilon_z=l  \epsilon_x$.
Therefore, 
combining together (\ref{actionregionI}) and (\ref{actionregionII}),
we finally obtain the Weyl anomaly (\ref{HoloWeylanomaly})
\begin{eqnarray}\label{HoloWeylanomaly}
\mathcal{A}=\int_{\partial M} \sqrt{h} \;  \frac{\alpha_5}{2} F_{na}F^{na} d^4 x .
\end{eqnarray}
Hence we obtain the universal relation (\ref{5drelation}).

We remark that we have $F_{ab}=0$ for the solution (\ref{vector5d}).
To get the information of central charge $b_4$ in
(\ref{5dWeylanomaly}), we need to consider solutions with $y^a$
dependence. We leave this problem to future study.

%%%%%%%%%%%%%%%%%%%%%%%%%%%%%%%%%%%
\section{Conclusions and Discussions}
%%%%%%%%%%%%%%%%%%%%%%%%%%%%%%%%%%%

In this paper, we have studied the anomalous current transport in holographic
BCFTs. We have verified that the holographic current is determined
universally by the central charge of the Weyl anomaly. To leading order of
nearness to the boundary, the current independent
of boundary conditions in four dimensions while it depends on boundary
conditions in higher dimensions. The holographic results obtained here support
the results obtained recently in \cite{Chu:2018ksb}.
We have also studied the back reaction of bulk
Maxwell's fields to AdS spacetime and obtain the holographic Weyl
anomaly for 5d BCFTs. It will be interesting to study Schwinger effect
\cite{schwinger} near the boundary. With the help of the boundary, the
magnetic field can separate the virtual particle pairs and turn them
into real particles. Thus, it is expected that constant magnetic field
can produce non-trivial Schwinger effect in the vicinity of the
boundary. We leave this problem for future study.

 \section*{Acknowledgements}
 We thank Dimitrios Giataganas, Bei-Lok Hu, Ling-Yan Hung, Yan Liu, Jian-Xin Lu
 and Tadashi Takayanagi
for useful discussions and comments.
This work is supported in part by 
%the National Center of Theoretical Science
NCTS and the grant MOST
105-2811-M-007-021 of the Ministry of
Science and Technology of Taiwan.

%%%%%%%%%%%%%%%%%%%%%%%%%%%%%%%%%%%%%%

\newpage

%%%%%%%%%%%%%%%%%%%%%%%%%%%%%%%%%%%%%%

\end{document}